\documentclass[aps,twocolumn,showpacs,superscriptaddress,floatfix]{revtex4-1}
\usepackage{amsfonts,hyperref}

\usepackage[draft]{todonotes}

\usepackage{graphicx}
\usepackage{amsmath}
\usepackage{amssymb}

\newcommand{\abs}[1]{\mbox{$\left| #1 \right|$}}

\begin{document}
\title{Period 4 stripe in the extended two-dimensional Hubbard model}

\author{Boris \surname{Ponsioen}} \affiliation{Institute for Theoretical Physics Amsterdam and Delta Institute for Theoretical Physics, University of Amsterdam, Science Park 904, 1098 XH Amsterdam, The Netherlands}
\author{Sangwoo S.~\surname{Chung}} \affiliation{Institute for Theoretical Physics Amsterdam and Delta Institute for Theoretical Physics, University of Amsterdam, Science Park 904, 1098 XH Amsterdam, The Netherlands}
\author{Philippe \surname{Corboz}}  \affiliation{Institute for Theoretical Physics Amsterdam and Delta Institute for Theoretical Physics, University of Amsterdam, Science Park 904, 1098 XH Amsterdam, The Netherlands}

\date{\today}

\begin{abstract}
We study the competition between stripe states with different periods and a uniform $d$-wave superconducting state in the extended 2D Hubbard model at 1/8 hole doping using infinite projected entangled-pair states (iPEPS). With increasing strength of negative next-nearest neighbor hopping~$t'$, the preferred period of the stripe decreases. For the values of $t'$ predicted for cuprate high-$T_c$ superconductors, we find stripes with a period 4 in the charge order, in agreement with experiments. Superconductivity in the period 4 stripe is suppressed at $1/8$ doping. Only at larger doping, $0.18 \lesssim \delta < 0.25$, the period 4 stripe exhibits coexisting $d$-wave superconducting order. The uniform $d$-wave state is only favored for sufficiently large positive $t'$.

\end{abstract}

\maketitle

\section{Introduction}

%%%%%%%%%%%%%%%%%%%%%%%%%%%%%%%%%%%%%%%%%%%%%%%%%%%%%%%%%%%%%%%%%%%%%%%%%%%%%%%%
% INTRO
%%%%%%%%%%%%%%%%%%%%%%%%%%%%%%%%%%%%%%%%%%%%%%%%%%%%%%%%%%%%%%%%%%%%%%%%%%%%%%%%

While enormous experimental and theoretical effort has been invested to understand the physics of  high-$T_c$ superconductivity in copper-oxide materials (or cuprates) \cite{Bednorz1986}, a full understanding of their phase diagram is still lacking \cite{Lee2006,Keimer2015,Robinson2019}.  
In the underdoped region of various cuprate materials, signatures of multiple broken symmetries have been inferred from experiments -- from neutron scattering \cite{tranquada95, Tranquada1996, yamada98, Tranquada2004,birgeneau06}, x-ray resonant scattering \cite{Abbamonte2005, Ghiringhelli2012,comin16,miao17}, nuclear magnetic resonance \cite{Hunt1999,wu11,wu2015, Imai2017} to scanning tunneling microscopy \cite{hoffman02,howald03,Wise2008,daSilvaNeto2014,mesaros16} -- collectively providing evidence for simultaneous charge and spin modulated states, coexisting or competing with superconductivity, called stripes \cite{Zaanen1989, Poilblanc1989, Schulz1989, Machida1989, emery99, kivelson03,Berg2007, ogata08,vojta09,fradkin12,fradkin15,kloss16,agterberg19}. %. with coexisting $d$-wave 

Meanwhile, the two-dimensional (2D) Hubbard model~\cite{Hubbard1963} --- which has been suggested as the most elementary microscopic model that may reproduce the essential features of the cuprates' phase diagram --- has been a prominent subject of intense theoretical and numerical investigations. Much attention has been given especially on the underdoped region in the strongly correlated regime, where several low-energy states are very closely competing, including a uniform $d$-wave superconducting (SC) state~\cite{giamarchi91,dagotto94,halboth00, maier05, capone06, aichhorn07, eichenberger07, tocchio08, kancharla08,yokoyama12, sordi12, gull12, gull13,chen13,kaczmarczyk13,otsuki14,deng15,tocchio16} and various stripe states~\cite{Zaanen1989, Poilblanc1989, Schulz1989, Machida1989,white03,hager05,chang10,Zheng2017,zhao17,vanhala18,Ido18,huang18, darmawan18,tocchio19} with or without coexisting superconducting order. 
A similar competition can also be found for the $t-J$ model --- the effective model in the strong coupling limit~\cite{white98_tJ,himeda02,ogata03,lugas06,raczkowski07,yang09,chou10,hu12,corboz14_tJ,dodaro17,Jiang2018a}. 
Computing the phase diagram of these models is a major challenge, as Quantum Monte Carlo suffers from the negative sign problem and the candidate ground states are very close in energies, calling for a high numerical precision; see Ref.~\cite{leblanc15} for a recent benchmark paper.

Recently, a consensus on the ground state at hole doping $\delta=1/8$ in the strongly correlated regime has been reached on the basis of several state-of-the-art numerical methods~\cite{Zheng2017} --- density matrix renomalization group (DMRG) \cite{White1992}, auxiliary field quantum Monte Carlo (AFQMC) \cite{Blankenbecler1981, Sugiyama1986}, density matrix embedding theory (DMET) \cite{Knizia2012}, and infinite projected entangled paired states (iPEPS) \cite{Verstraete2004, Jordan2008} --- that the ground state is a stripe state with a period of 8 sites in the charge order without coexisting $d$-wave SC order, while stripes with periods 5-7 being energetically very close. This result was also confirmed by variational Monte Carlo (VMC)~\cite{Ido18, tocchio19}.
However, the period 4 stripe typically observed in experiments at 1/8 doping~\cite{tranquada95,tranquada97,mesaros16}  was found to be higher in energy~\cite{Zheng2017}. Consequently, it is evident that a more realistic model of the cuprates than the most basic Hubbard model needs to be considered.

In this paper, using state-of-the-art iPEPS simulations, we show that a period 4 stripe ground state is obtained in an extended Hubbard model including a realistic next-nearest neighbor (NNN) hopping at $\delta=1/8$ doping.  Our systematic study demonstrates that: (\emph{i}) the ground-state stripe period decreases  with the magnitude of the negative NNN hopping amplitude $t'$; (\emph{ii}) there exists a large region of $t'$ in which the period 4 stripe is stabilized, including realistic values for $t'$ predicted for the cuprates~\cite{hirayama18,hirayama19}; and finally that (\emph{iii}) the $d$-wave SC order does not coexist with the period 4 stripe at $\delta=1/8$ doping, but only at larger doping. Our observations of a decrease in the stripe period with negative $t'$ are also in agreement with recent works based on DMRG on width-4 cylinders~\cite{huang18,Jiang1424} and VMC~\cite{Ido18}. 

%%%%%%%%%%%%%%%%%%%%%%%%%%%%%%%%%%%%%%%%%%%%%%%%%%%%%%%%%%%%%%%%%%%%%%%%%%%%%%%%
% MODEL 
%%%%%%%%%%%%%%%%%%%%%%%%%%%%%%%%%%%%%%%%%%%%%%%%%%%%%%%%%%%%%%%%%%%%%%%%%%%%%%%%

%\emph{Model.} --- 
\section{Model}
We consider an extended Hubbard model on a two-dimensional (2D) square lattice given by the Hamiltonian
\begin{eqnarray}
H&=&-t\sum_{\left\langle i,j\right\rangle ,\sigma} \hat c_{i\sigma}^{\dagger}  \hat c_{j\sigma} + h.c. \nonumber \\ 
     &&-t'\sum_{\left\langle \left\langle i,j\right\rangle \right\rangle, \sigma}  \hat c_{i\sigma}^{\dagger}  \hat c_{j\sigma} + h.c. +U\sum_{i}  \hat n_{i\uparrow}  \hat n_{i\downarrow},
\end{eqnarray}
where $t$ and $t'$ are nearest neighbor and NNN hopping amplitudes, $U$$>$0 is the on-site Coulomb repulsion, and $ \hat c_{i\sigma}$ $( \hat c_{i\sigma}^{\dagger})$ is the annihilation (creation) operator for an electron of spin $\sigma$ on site $i$, and $\hat n_{i\sigma} = \hat c_{i\sigma}^{\dagger} \hat c_{i\sigma}$.  We focus on $U/t=10$, which has been predicted as a realistic value of the cuprates~\cite{hirayama18}, and on $\delta$=1/8 throughout this work, unless stated otherwise.

%%%%%%%%%%%%%%%%%%%%%%%%%%%%%%%%%%%%%%%%%%%%%%%%%%%%%%%%%%%%%%%%%%%%%%%%%%%%%%%%
% METHOD
%%%%%%%%%%%%%%%%%%%%%%%%%%%%%%%%%%%%%%%%%%%%%%%%%%%%%%%%%%%%%%%%%%%%%%%%%%%%%%%%

%Corboz2010b

%\emph{Method.} --- 
\section{Method}
For our simulations, we apply the fermionic implementation \cite{corboz2010,kraus2010,Barthel2009} of iPEPS --- a tensor network variational ansatz \cite{Verstraete2004, nishio2004, Jordan2008, Verstraete2009} for 2D lattice systems in the thermodynamic limit --- which has gained recognition as a reliable and versatile numerical technique for 2D strongly correlated systems (see e.g. Refs.~\cite{Corboz2011su4, Zhao2012, Corboz2014b,corboz14_tJ,niesen17,Liao2017,Haghshenas2018,Chen2018,Jahromi2018,chung19} and references therein).  
The ansatz consists of a supercell of tensors that is periodically repeated on the lattice, with one tensor per lattice site. Each tensor has one physical index carrying the local Hilbert space of a lattice site and four auxiliary indices connecting neighboring tensors. The accuracy of the ansatz can be systematically controlled by the bond dimension $D$ of the auxiliary indices. Translationally invariant states can be represented by an iPEPS with a single-tensor supercell. If translational symmetry is spontaneously broken, a larger supercell compatible with the symmetry breaking pattern is required.

For technical details on iPEPS we refer to Refs.~\cite{corboz2010,phien15}. For the experts, we note that the optimization of the iPEPS wave function (i.e. finding the optimal variational parameters) is done using an imaginary time evolution based on a 3-site cluster update~\cite{wang11b,niesen18}, in which the 2D wave function is only taken into account in an effective way during the optimization~\cite{Jiang2008}.  This allows us to reach large bond dimensions of up to $D=18$ even in the presence of a next-nearest neighbor hopping. Observables are computed by contracting the two-dimensional tensor network using the corner transfer matrix method~\cite{nishino1996, Orus2009} generalized to arbitrary supercell sizes~\cite{Corboz2011,corboz14_tJ}. To increase the efficiency, we exploit abelian $U(1)$ symmetries of the model~\cite{singh2010,bauer2011}.

%%%%%%%%%%%%%%%%%%%%%%%%%%%
\begin{figure}[]
  \centering
  \includegraphics[width=\linewidth]{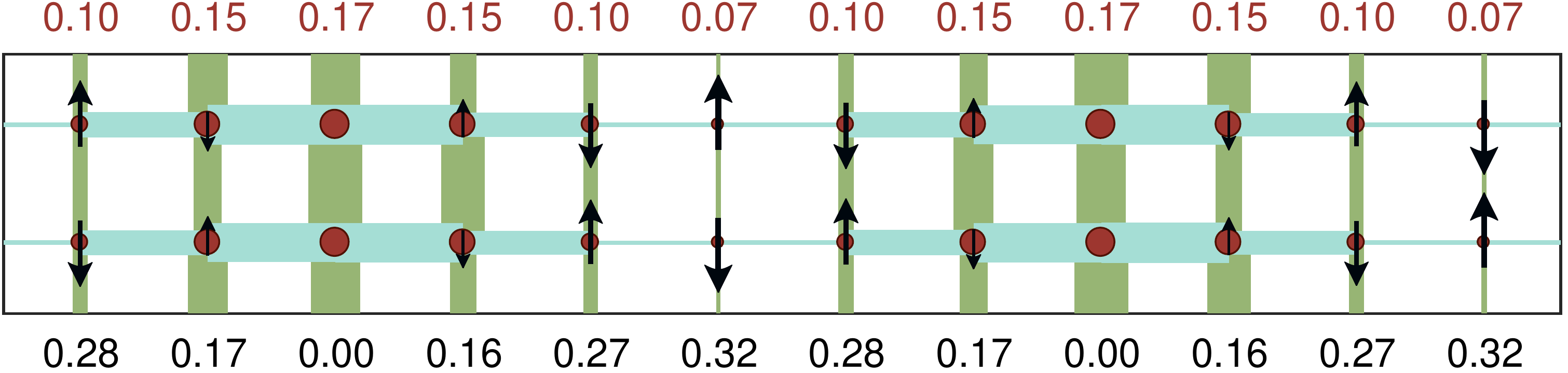}
  \caption{Example stripe with a period 6 in the charge order and period 12 in the spin order obtained with iPEPS using a $12\times2$ supercell, for $U/t=10$, $t'/t=0$, $D=12$. The sizes of the red discs and the black arrows scale with the local hole density and the local magnetic moment, respectively, with average values indicated in the top and bottom rows. The width of a bond is proportional to the local singlet pairing amplitude on the corresponding bond, with different signs in x- and y- direction represented by the two different colors.   }
  \label{fig:w6}
\end{figure}
%%%%%%%%%%%%%%%%%%%%%%%%%%%

To identify the ground state, energies of various competing low-energy states obtained with different supercell sizes are compared, including a uniform $d$-wave  SC state   with coexisting antiferromagnetic (AF) order, obtained in a $2\times2$ supercell, and stripe states with different periods in the charge order. Stripes with odd periods 3, 5, 7 (designated W3, W5, W7) are described by 3$\times$2, 5$\times$2, and 7$\times$2 supercells, respectively. For even period stripes (W4, W6, W8), 8$\times$2, 12$\times$2 and 16$\times$2 supercells are used, since in these cases the period of the spin order is twice the period of the charge order due to the $\pi$-phase shift in the AF order across the sites with maximal hole density.

An example stripe with period 6 (W6) is shown in Fig.~\ref{fig:w6}, which visualizes the local hole density $\delta_i = 1 - n_i$ and the local magnetic moment $s^z_i = \frac12 \langle \hat n_{i\uparrow} - \hat n_{i\downarrow} \rangle$ on each site~$i$, and the singlet pairing amplitude $\Delta^s_{ij}=\langle \hat c_{i\uparrow} \hat c_{j\downarrow} - \hat c_{j\uparrow} \hat c_{i\downarrow}  \rangle / \sqrt{2}$ between neighboring sites $i$ and $j$ within the supercell.

%%%%%%%%%%%%%%%%%%%%%%%%%%%%%%%%%%%%%%%%%%%%%%%%%%%%%%%%%%%%%%%%%%%%%
% RESULTS
%%%%%%%%%%%%%%%%%%%%%%%%%%%%%%%%%%%%%%%%%%%%%%%%%%%%%%%%%%%%%%%%%%%%%

\section{Results}

\subsection{Shift of stripe period as a function of $t'/t$}%
\label{sec:results}
Previously, in Ref.~\cite{Zheng2017}, it was found that for $U/t=8$, $\delta=1/8$, and $t'=0$ stripe states have lower energies than the uniform $d$-wave SC state. A close competition in energies between stripes with periods 5 to 8 was found, with a slight preference towards the period 8 stripe (W8), while the experimentally observed period 4 stripe (W4) was clearly higher in energy. In the following, we study the effect of an additional next-nearest neighbor hopping on the preferred stripe period, using iPEPS simulations for a fixed bond dimension $D=12$.

%%%%%%%%%%%%%%%%%%%%%%%%%%%
\begin{figure}[]
  \centering
  \includegraphics[width=\linewidth]{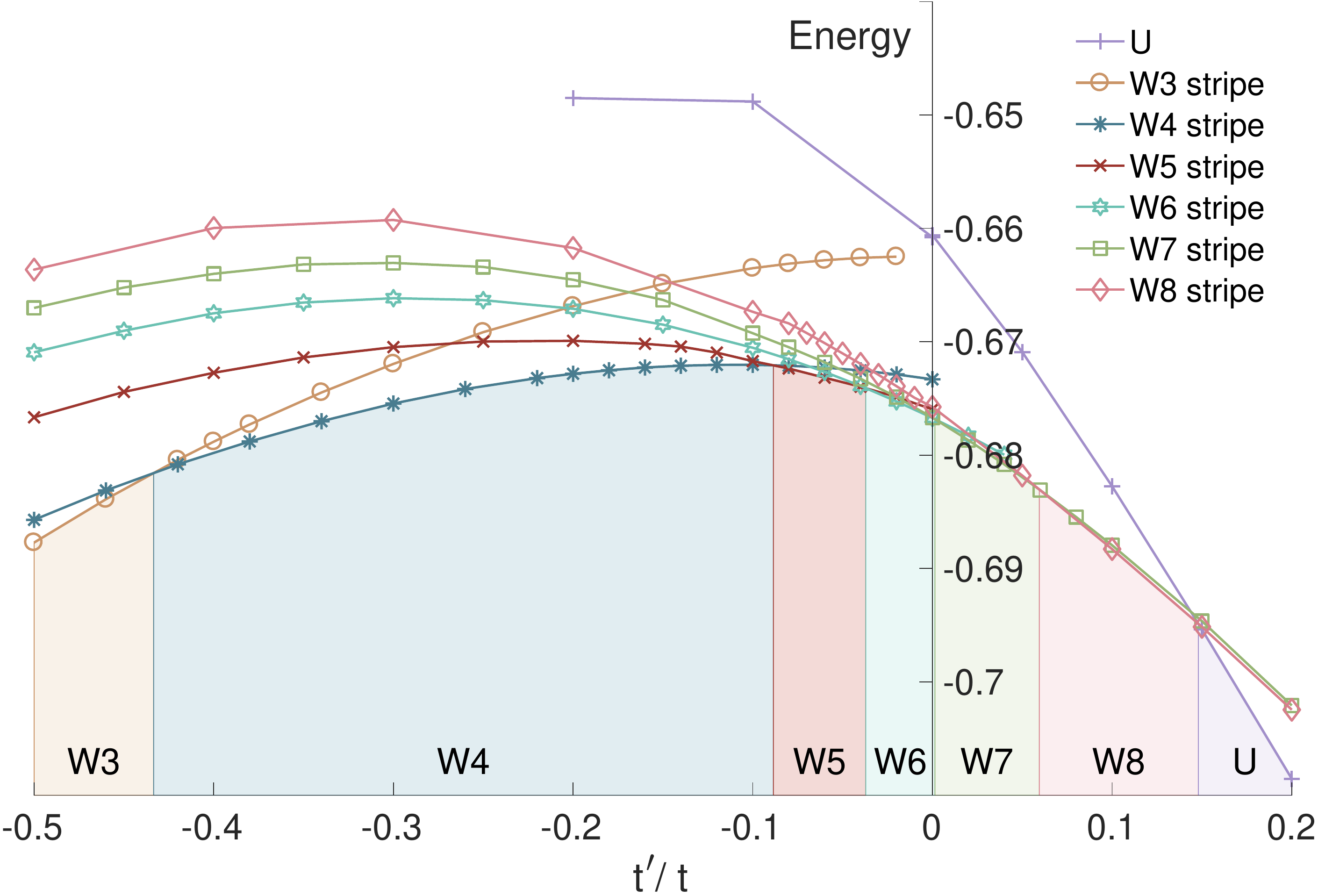}
  \caption{Phase diagram for $U/t=10$,  $\delta\!=\!1/8$, for a fixed bond dimension $D=12$ as a function of the next-nearest neighbor hopping strength~$t'/t$. Each region in the phase diagram is defined by the corresponding lowest energy state, including stripe states with periods between 3 and 8 (W3-W8) or a uniform $d$-wave SC state (U).}
  \label{fig:phasediagram} 
\end{figure}
%%%%%%%%%%%%%%%%%%%%%%%%%%%

In Fig.~\ref{fig:phasediagram}, we present the energies of the competing states and the resulting phase diagram for $U/t=10$, $\delta\!=\!1/8$ as a function of $t'/t$. In agreement with the previous results for $U/t=8$~\cite{Zheng2017}, we find several closely competing stripe states with periods 5-8 around $t'/t\sim0$. For $t'/t=0$, we find a slight preference towards W6/W7 stripes instead of the W8 stripe~\cite{Zheng2017}.  This can be attributed to the larger value of $U/t$ used here, leading to favor smaller periods~\cite{Zheng2017}, and also to the finite bond dimension $D=12$, which may lead to slight relative shifts in the energies.

For negative $t'$, as we increase $\abs{t'/t}$, we observe a gradual shift of the preferred stripe period to smaller periods. In particular, the period 4 (W4) stripe becomes the ground state in a large region of the phase diagram, between $-0.43 \lesssim t'/t \lesssim -0.09$ for $D=12$. We will discuss the $D$-dependence of the W4 phase boundaries in the next section. We note that a shift to smaller stripe periods was also found recently with DMRG on width-4 cylinders~\cite{huang18,Jiang1424} and VMC~\cite{Ido18}.
%\todo[inline]{Note: same as the introduction: this phrasing seems to suggest VMC on w4 cylinders - shall we change it?}

It is interesting to observe that while various stripe periods are very close in energies for $t'/t=0$, the competition between several states becomes less strong for negative $t'/t$, especially deep in the W4 phase. In this respect, the extended Hubbard model with a negative $t'/t$ is less challenging to study than the Hubbard model without a $t'$.

Intuitively, the shift in preferred stripe period can be understood by considering the effective extended \emph{t--J} model, obtained using a second order perturbative expansion in $t/U$ ($t'/U$), where neighboring (NNN) spins interact via an AF Heisenberg interaction with amplitude  $J = 4 {t}^2/U$ ($J' = 4 {t'}^2/U$). The strong AF order favored by the nearest-neighbor term gets frustrated by the NNN term, such that  extended AF regions in long-period stripes become energetically unfavorable with increasing $J'$, and stripes with shorter periods eventually attain lower energies.
 In order to get more insights into the energetics of the stripes, we present detailed data and a discussion of different local energy contributions of two example stripes in appendix \ref{sec:app}.

Finally we note that for sufficiently large positive $t'/t\gtrsim 0.148$, the uniform state becomes energetically favorable over stripes, in agreement with previous findings~\cite{huang18}. The intuitive reason is that with a positive $t'/t$, holes can delocalize along the diagonals without frustrating the antiferromagnetic order in the uniform state; whereas in a stripe state, holes on the domain wall moving diagonally frustrate the antiferromagnetic order, such that stripes become disfavored for sufficiently large~$t'/t$.

\subsection{Stability and extension of the W4 stripe phase}%
\begin{figure}[]
  \centering
  \includegraphics[width=1\linewidth]{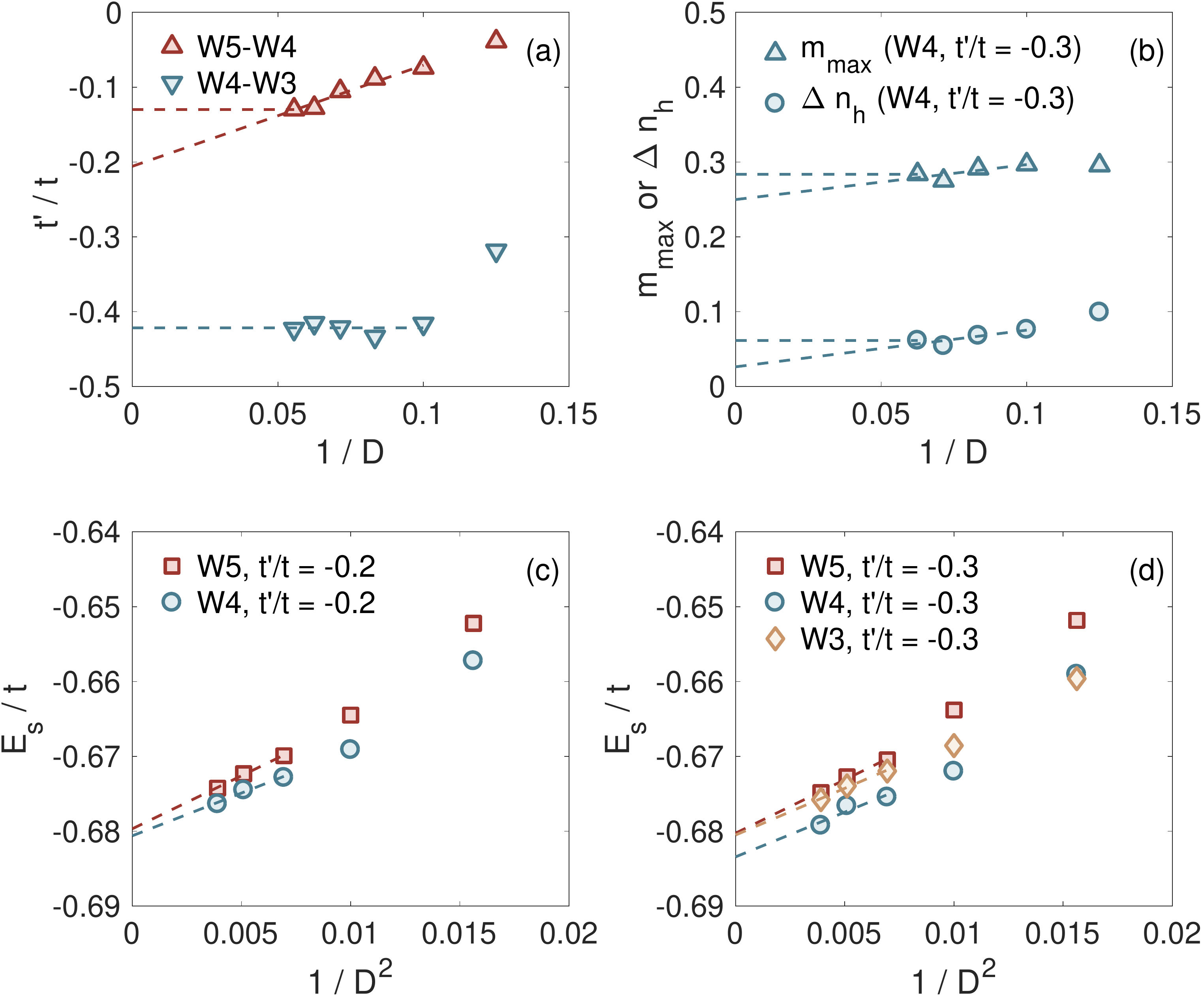}
  \caption{(a) Phase boundaries of the period 4 (W4) stripe phase with its adjacent W5 and W3 stripe phases  as a function of the inverse bond dimension. (b) Maximal local magnetic moment and amplitude of the hole density  modulation of the W4 stripe for $t'/t=-0.3$. Bottom: energy per site of the W3, W4 and W5 stripes as a function of the inverse squared bond dimension at $t'/t=-0.2$ (c) and $t'/t=-0.3$ (d). Dashed lines are a guide to the eye. }
  \label{fig:w4_w5_vs_E}
\end{figure}

Having studied the qualitative features of the phase diagram at fixed $D=12$, in this section we present a systematic study of the stability of the W4 stripe phase as a function of the bond dimension. The $D$-dependence of the phase boundaries of the W4 phase and its adjoining phases is shown in Fig.~\ref{fig:w4_w5_vs_E}. While we observe a shift of the W4/W5 phase boundary for $D \leq 16$, the change between $D=16$ and $D=18$ is small. Thus we expect that a linear extrapolation in $1/D$ of the phase boundary provides a (conservative) lower bound of the phase transition ($t'/t=-0.21$), and the transition value at the largest $D=18$, $t'/t=-0.12$, an upper bound. For the W3/W4 phase boundary we only find a weak and non-monotonous $D$ dependence for $D\ge10$ with a transition value around $t'/t = -0.423(10)$. 

The stability of the W4 stripe is further supported by the results in Fig.~\ref{fig:w4_w5_vs_E}(c) and (d), where we compare the energies of the competing stripes for $t'/t=-0.2$ and $-0.3$, and show that the W4 stripe has the lowest energy even in the infinite $D$ limit~\footnote{Since the energy typically converges faster than linearly in $1/D$, we plot the energy as a function of $1/D^2$ instead for the comparison of the states.}. In Fig.~\ref{fig:w4_w5_vs_E}(b), we present data of the maximum of the local magnetic moment in the stripe $m_{max}$ and the amplitude of the hole density modulation $\Delta n_h$, given by the difference between the maximal and minimal hole density within the supercell. We find that both orders remain finite in the infinite $D$ limit, which indicates that both spin and charge stripe order are stable. This is in contrast to the DMRG results on width-4 cylinders~\cite{Jiang1424} where the spin order was found to be suppressed for $t'/t=-0.25$.

Thus from these results, we conclude that the W4 stripe is the ground state in a large region between $0.16(4)< -t'/t < 0.423(10)$, which includes realistic values predicted for cuprate materials~\cite{hirayama18,hirayama19}.

\subsection{Pairing in the W4 stripe state}
Previously in Ref.~\cite{Zheng2017}, it was found for $t'=0$ and $\delta=1/8$ that W5-W7 stripes exhibit coexisting $d$-wave SC order (see also Fig.~\ref{fig:w6}), whereas in the W8 stripe, which has a filling of one hole per unit length of a stripe (i.e. $\rho_l=1$), the pairing is entirely suppressed. Here we find qualitatively similar results also at finite $t'$ for these stripes (see e.g. the W6 stripe in Fig.~\ref{fig:w6} with coexisting $d$-wave pairing).

\begin{figure}[t]
\begin{center}
\includegraphics[width=1\columnwidth]{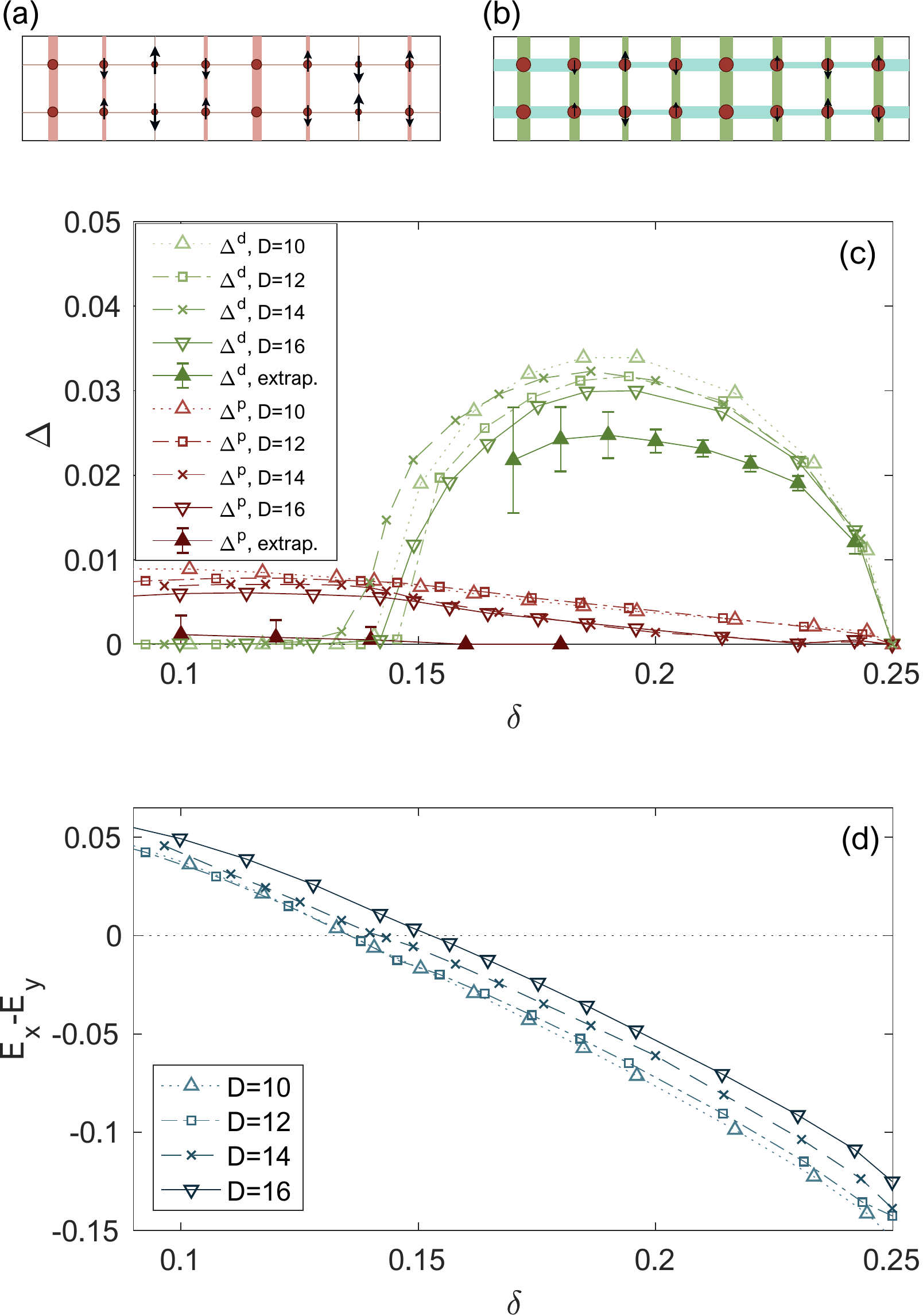} 
\caption{(Color online) (a) W4 stripe at $1/8$ doping and $t'/t=-0.2$. The colored red bonds show the strength of the nearest-neighbor triplet pairing amplitude $\Delta^t_{ij}$ along the stripe, obtained at finite $D=12$. (b) W4 stripe at larger doping $\delta \sim 0.2$ (same parameters as in (a)) coexisting with $d$-wave pairing. (c) Mean $d$-wave pairing (green data) and $p$-wave pairing (red data) as a function doping $\delta$ for different bond dimensions $D$. The extrapolated value (full symbols) of the $p$-wave pairing vanishes in the entire doping range whereas the extrapolated $d$-wave pairing remains finite at large doping. (d) Difference of the energy per bond in $x$ and $y$ directions.}
\label{fig:W4stripe}
\end{center}
\end{figure}

The dominant pairing correlations of the W4 stripe at 1/8 doping are, however, very different from the ones of the W5-W7 stripes: the mean $d$-wave pairing per site 
 vanishes and instead we find, for finite $D$, a $p$-wave pairing with finite nearest-neighbor triplet pairing amplitude $\Delta^t_{ij}=\langle \hat c_{i\uparrow} \hat c_{j\downarrow} + \hat c_{j\uparrow} \hat c_{i\downarrow}  \rangle / \sqrt{2}$ in the longitudinal direction of the stripe (however, as we will show below, the triplet pairing amplitude most likely vanishes in the infinite $D$ limit). The finite-$D$ pairing pattern of an example W4 stripe for $\delta \sim 1/8$ is shown in Fig.~\ref{fig:W4stripe}(a), where the thickness of the bonds scales with the magnitude of~$\Delta^t_{ij}$.

In order to get more insights into the pairing order in the W4 stripe, we study its properties and $D$-dependence over an extended doping range~\footnote{We stress that the W4 stripe may not be the ground state over the entire doping range presented here. Here we consider dopings away from 1/8 in order to better understand the difference in the superconducting order between the W4 stripe and the W5-W7 stripes at 1/8 doping, and not to make a statement about the ground state away from 1/8 doping.}.
%
%~\footnote{We note that the W4 stripe may not be the ground state over the entire doping range presented here, but we can nevertheless study its doping dependence here to get insights into the pairing properties}. 
%
In Fig.~\ref{fig:W4stripe}(c) we present results for the mean $d$-wave ($p$-wave) pairing amplitude $\Delta^d$ ($\Delta^p$), given by the site-averaged $\Delta^s_{ij}$ ($\Delta^t_{ij}$) with different (same) phase factors in x- and y-direction. These results reveal
two different regions: a region with $d$-wave pairing at large doping, $\delta \gtrsim 0.14$, and a region with predominant $p$-wave pairing for $\delta \lesssim 0.14$, which includes $\delta=1/8$. An example of a state with $d$-wave pairing for $\delta\sim0.2$ is shown in Fig.~\ref{fig:W4stripe}(b).  
While the $d$-wave pairing  remains finite when extrapolating the data in $1/D$, the $p$-wave pairing is more strongly suppressed with increasing $D$, with the extrapolated value being compatible with a vanishing pairing amplitude. This implies that the triplet pairing observed at finite $D$ only reflects the short-range pairing correlations, but in the exact infinite $D$ limit, there is no true long-range  SC order for  $\delta \lesssim 0.14$.

Thus, these results indicate that the $W4$ stripe undergoes a phase transition from a non-superconducting state to a state with dominant $d$-wave pairing around a doping of $\delta \sim 0.14$. Interestingly, these two regions are also characterized by different  energy strengths in transverse ($E_x$) and longitudinal ($E_y$) direction; see Fig.~\ref{fig:W4stripe}(d). In the short-range $p$-wave region at low doping, the longitudinal energy contributions are stronger than the transverse ones $E_y<E_x$, reminiscent of weakly coupled chains (in which one would expect power-law decaying pairing correlations), whereas in the $d$-wave region at large doping, we find the opposite, i.e. $E_x<E_y$. This qualitative change in the  energy contributions in the two spatial directions was also previously observed in the $t$-$J$ model~\cite{corboz14_tJ}.

We note that a similar qualitative change in the pairing can also be observed in the other stripes, but with the transition value $\delta_c$ between the two regions shifted to smaller dopings, e.g. $\delta_c \sim 0.115$ for the W5 stripe (i.e. 1/8 doping is within the $d$-wave region of the W5 stripe). Interestingly, when comparing the hole density per unit length of the stripe, $\rho_l = \delta \cdot W$, with $W$ the width of the stripe, we find a similar transition value $\rho_l \sim 0.57$ for both the W4 and W5 stripe.

%We note that the absence of superconductivity in the W4 stripe at $1/8$ doping is also compatible with the VMC results in Ref.~\cite{Ido18}. However, in contrast to our results, the stripes found in VMC exhibit strongly suppressed $d$-wave SC order over the entire doping range.

We note that the absence of superconductivity in the W4 stripe at $1/8$ doping is also compatible with the VMC results in Ref.~\cite{Ido18}. However, in contrast to the VMC results we find a finite SC order in the W5 - W7 stripes at 1/8 doping (which are the ground states for smaller negative $t'/t$).

%
%%%%%%%%%%%%%%%%%%%%%%%%%%%%%%%%%%%%%%%
%%%%%%%%%%%%%%%%%%%%%%%%%%%%%%%%%%%%%%%

%%%%%%%%%%%%%%%%%%%%%%%%%%%%%%%%%%%%%%%%%%%%%%%%%%%%%%%%%%%%%%%%%%%%%
% CONCLUSION
%%%%%%%%%%%%%%%%%%%%%%%%%%%%%%%%%%%%%%%%%%%%%%%%%%%%%%%%%%%%%%%%%%%%%

\section{Summary and conclusion}
We have studied the ground state phases of an extended Hubbard model with a next-nearest neighbor hopping $t'$ using iPEPS, focusing on $U/t=10$ and $\delta=1/8$ doping. The preferred stripe period decreases with increasing strength of the negative $t'$. In particular, in agreement with experiments~\cite{tranquada95,mesaros16}, we have found a period 4 stripe as the ground state in an extended parameter region, $0.16(4)< -t'/t < 0.423(10)$,  including also realistic values predicted for cuprate materials~\cite{hirayama18,hirayama19}. 

%mesaros16

Superconductivity in the period 4 stripe at 1/8 doping is suppressed, where the dominant nearest-neighbor pairing correlations are not in the singlet but in the triplet channel (without long-range order in the infinite $D$ limit).  However, at larger doping of $\delta \gtrsim 0.14$  (or hole density per unit length $\rho_l \gtrsim 0.57$), the period 4 stripe exhibits coexisting $d$-wave SC order, showing that the pairing nature of the stripe depends on doping. 

For a realistic negative $t'/t$, the competition among the stripe and uniform states turns out to be weaker than for the simplest Hubbard model without a $t'$, and thus from a numerical point of view the former model is less challenging to study.  The same may be true for even more realistic multi-band Hubbard models, which can also be efficiently studied using iPEPS, offering a promising perspective to get further insights into the various competing phases observed in the cuprates.

%%%%%%%%%%%%%%%%%%%%%%%%%%%%%%%%%%%%%%%%%%%%%%%%%%%%%%%%%%%%%%%%%%%%%%%%%%%%%%%%
% ACKNOWLEDGMENTS
%%%%%%%%%%%%%%%%%%%%%%%%%%%%%%%%%%%%%%%%%%%%%%%%%%%%%%%%%%%%%%%%%%%%%%%%%%%%%%%%
\begin{acknowledgments}
We acknowledge helpful discussions with T.~M.~Rice and N.~J.~Robinson. This project has received funding from the European Research Council (ERC) under the European Union's Horizon 2020 research and innovation programme (grant agreement No 677061). This work is part of the D-ITP consortium, a program of the Netherlands Organization for Scientific Research (NWO) that is funded by the Dutch Ministry of Education, Culture and Science~(OCW). 
\end{acknowledgments}

B.P.~and S.S.C.~contributed equally to this work.

%%%%%%%%%%%%%%%%%%%%%%%%%%%%%%%%%%%%%%%%%%%%%%%%%%%%%%%%%%%%%%%%%%%%%
% APPENDIX
%%%%%%%%%%%%%%%%%%%%%%%%%%%%%%%%%%%%%%%%%%%%%%%%%%%%%%%%%%%%%%%%%%%%%

\appendix

\section{Comparison of local energy contributions of the W7 and W4 stripes}
\label{sec:app}
\begin{figure}[t!]
\begin{center}
\includegraphics[width=1\columnwidth]{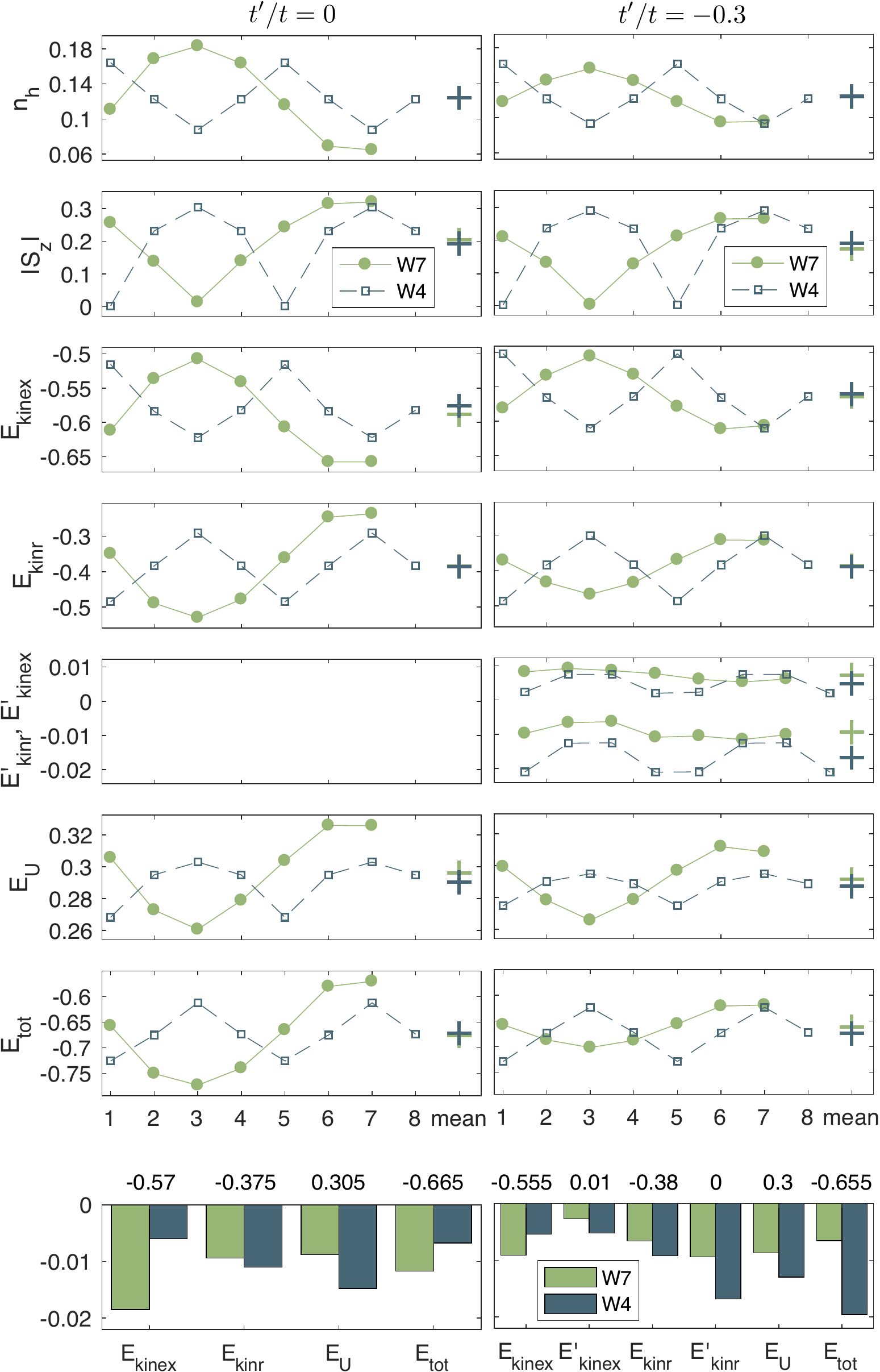} 
\caption{(Color online) Local hole density, local magnetic moment,  and various local energy contributions (cf. text) of the W7 and W4 stripes for $t'/t=0$ (left panels) and $t'/t=-0.3$ (right panels), for $U/t=10$, $\delta=1/8$, $D=12$ on each site (or bond) of the two stripes.  The bottom panels show the average energy contributions. Each pair of bars has been shifted by the value indicated in the top row. All energies are in units of $t$. }
\label{fig:E}
\end{center}
\end{figure}

In this appendix, we present a detailed comparison of various local energy contributions of two stripes, W7 and W4, for two different values of the NNN hopping, $t'/t=0$ and $t'/t=-0.3$, for $\delta=1/8$. We split the local kinetic energy into two parts, $E_{kin}=E_{kinex} +E_{kinr}$, where the first part, $E_{kinex}$, is the one relevant for magnetic superexchange processes in the Heisenberg limit,   i.e. matrix elements between single occupied sites with opposite spin \{ $|\uparrow,\downarrow\rangle$, $|\downarrow,\uparrow\rangle$\} and doubly occupied sites with a hole \{$|\uparrow\downarrow,0\rangle$, $|0,\uparrow\downarrow\rangle$\}. The second part, $E_{kinr}$ contains all the remaining kinetic energy contributions, i.e. matrix elements between \{$|\uparrow,0\rangle$,  $|\downarrow,0\rangle$, $|\uparrow\downarrow,\downarrow\rangle$,  $|\uparrow\downarrow,\uparrow\rangle$\} and \{$|0, \uparrow\rangle$,  $|0,\downarrow\rangle$, $|\downarrow,\uparrow\downarrow\rangle$,  $|\uparrow,\uparrow\downarrow\rangle$\}. The column-averaged energy contributions together with the local hole density $n_h$ and magnitude of the local magnetic moment $|s_z|$ are presented in Fig.~\ref{fig:E}, where $E_U$ denotes the interaction term, $E_{tot}$ the total energy, and $E'_{kinex}$ and $E'_{kinr}$ the NNN kinetic contributions. A comparison of the average contributions of the two stripes is presented in the bar plots in the bottom panels.

%%%%%%%%%%%%%%%%%%%%%%%%%%%%%%%%%%%%%%%
%%%%%%%%%%%%%%%%%%%%%%%%%%%%%%%%%%%%%%%

First we observe, as expected, that $E_{kinex}$  is lowest in the region with strong AF order (i.e. the region with small hole density),  whereas the other part  $E_{kinr}$ is lowest around the maximum hole density, where the holes can fluctuate  without frustrating the AF order, thanks to the $\pi$ phase shift in the AF order. (The $E_{kinex}$ energy is partly compensated by a positive $E_{U}$ contribution of having doubly occupied sites in the exchange processes.) Thus there is energetic tradeoff between having large AF regions (as can be found in long period stripes) yielding a low $E_{kinex}$, and introducing more $\pi$-phase shifts (i.e. shorter period stripes) enhancing the $E_{kinr}$ part.

For $t'/t=0$, the W7 stripe has a lower total energy than the W4 stripe, mostly thanks to a lower $E_{kinex}$. The W4 stripe has  a lower $E_{kinr}$ than the W7 stripe, but the difference in $E_{kinex}$ is larger between the two (even when taking into account the additional energy cost $E_U$ to form doubly occupied sites) such that, overall, the W7 is energetically favored.

A finite $t'/t$   introduces an effective antiferromagnetic Heisenberg interaction along the diagonals, which frustrates the AF order such that extended AF regions become less favorable. This leads to a suppression of magnetic moments in the W7 stripe when going from  $t'/t=0$ to $t'/t=-0.3$, which is realized by increasing the hole density in the AF region and thereby reducing the charge and spin stripe amplitudes. The frustration along the diagonal is less severe in the W4 stripe with weaker AF regions. In addition, we observe that diagonal kinetic energies $E'_{kinr}$ are lower in regions with a larger difference in hole densities between neighboring rows.  Consequently, $E'_{kinr}$ is lower in the W4 stripe than in the W7 stripe, since the local hole density changes more rapidly in a shorter period stripe. Overall, this leads to a lower total energy of the W4 stripe compared to the W7 stripe for $t'/t=-0.3$.

\bibliographystyle{apsrev4-1}
\bibliography{refs,biblio}

\end{document}